\documentclass{desyproc}

\begin{document}
\title{Status of the ANAIS Dark Matter Project at the Canfranc Underground Laboratory}
\author{{\slshape J.~Amar\'{e}$^{1,2}$, S.~Cebri\'{a}n$^{1,2}$, C.~Cuesta$^{1,2}$\footnote{Present address: Department of Physics, Center for Experimental Nuclear Physics and Astrophysics, University of Washington, Seattle, WA, USA}, E.~Garc\'{i}a$^{1,2}$, M.~Mart\'{i}nez$^{1,2}$\footnote{Universit\`{a} di Roma La Sapienza, Piazzale Aldo Moro 5, 00185 Roma, Italy}, M.A.~Oliv\'{a}n$^{1,2}$\footnote{Attending speaker},
Y.~Ortigoza$^{1,2}$, A.~Ortiz de Sol\'{o}rzano$^{1,2}$, C.~Pobes$^{1,2}$\footnote{Instituto de Ciencia de Materiales de Arag\'{o}n, Universidad de Zaragoza-CSIC, Zaragoza, Spain}, J.~Puimed\'{o}n$^{1,2}$, M.L.~Sarsa$^{1,2}$, J.A.~Villar$^{1,2}$, and P.~Villar$^{1,2}$}\\[1ex]
$^1$Laboratorio de F\'{i}sica Nuclear y Astropart\'{i}culas, Universidad de Zaragoza, Calle Pedro Cerbuna 12, 50009 Zaragoza, Spain\\
$^2$Laboratorio Subterr\'{a}neo de Canfranc, Paseo de los Ayerbe s/n, 22880 Canfranc Estaci\'{o}n, Huesca, Spain}

\contribID{familyname\_firstname}

\confID{11832}  
\desyproc{DESY-PROC-2015-02}
\acronym{Patras 2015} 
\doi  

\maketitle

\begin{abstract}
The ANAIS experiment aims at the confirmation of the DAMA/LIBRA signal. A detailed analysis of two
NaI(Tl) crystals of 12.5~kg each grown by Alpha Spectra will be shown:
effective threshold at 1 keVee is at reach thanks to outstanding light collection and robust PMT
noise filtering protocols and the measured background is well
understood down to 3 keVee, having quantified K, U and
Th content and cosmogenic activation in the crystals. A new detector was
installed in Canfranc in March 2015 together with the two previous
modules and preliminary characterization results will be presented. Finally, the
status and expected sensitivity of the full experiment with 112~kg
will be reviewed.
\end{abstract}

\section{The ANAIS experiment}

The ANAIS (Annual modulation with NaI Scintillators) project is intended to search for dark matter annual modulation with ultrapure NaI(Tl) scintillators at the Canfranc Underground Laboratory (LSC) in Spain, in order to provide a model-independent confirmation of the signal reported by the DAMA/LIBRA collaboration~\cite{dama} using the same target and technique. Similar performance to DAMA/LIBRA detectors in terms of threshold and background are consequently mandatory.
The total active mass will be divided into modules, each consisting of a 12.5\,kg NaI(Tl) crystal encapsulated in copper and optically coupled to two photomultipliers (PMTs) working in coincidence. Nine modules in a 3$\times$3 matrix are expected to be set-up at LSC along 2016. The shielding for the experiment consists of 10\,cm of archaeological lead, 20\,cm of low activity lead, 40\,cm of neutron moderator, an anti-radon box, and an active muon veto system made up of plastic scintillators covering top and sides of the whole set-up. The experiment hut at the hall B of LSC (under 2450\,m.w.e.) is already operative and shielding materials, selected Hamamatsu R12669SEL2 PMTs and electronic chain components are ready. The main challenge of the project has been the achievement of the required crystal radiopurity. A 9.6\,kg NaI(Tl) crystal made by Saint-Gobain was first operated~\cite{ANAISbulk,ANAISbkg,anais40K,ANAISom} but disregarded due to an unacceptable K content. Two prototypes of 12.5\,kg mass, made by Alpha Spectra, Inc. Colorado with ultrapure NaI powder, took data at the LSC since December 2012 (ANAIS-25 set-up) and a new module also built by Alpha Spectra using improved protocols for detector production was added in March 2015 (ANAIS-37 set-up).

\section{The ANAIS-25 and ANAIS-37 set-ups}

The main goals for the ANAIS-25 set-up~\cite{ANAISricap13} were to
measure the crystal contamination, evaluate light
collection, fine tune the data acquisition and test the filtering
and analysis protocols. The two modules (named D0 and D1) are
cylindrical, 4.75$"$ diameter and 11.75$"$ length, with quartz
windows for PMTs coupling. A Mylar window in the lateral face allows
for low energy calibration. After testing other PMT model, Hamamatsu
R12669SEL2 units were used for both detectors. The modules were
shielded by 10\,cm of archaeological plus 20\,cm of low activity
lead at LSC. An impressive light collection at the level of
$\sim$15~phe/keV has been measured for these detectors~\cite{posterma}. Background contributions have been thoroughly
analyzed and Table~\ref{tab:cont} shows the results of the
activities determined for the main crystal contaminations: $^{40}$K
content has been measured performing coincidence analysis between
1461\,keV and 3.2\,keV energy depositions in different
detectors~\cite{anais40K} and the activities from $^{210}$Pb and
$^{232}$Th and $^{238}$U chains have been deduced by quantifying
Bi/Po sequences and the total alpha rate determined through pulse
shape analysis. The content of $^{40}$K, above the initial goal of
ANAIS (20\,ppb of K), is acceptable, $^{232}$Th and $^{238}$U
activities are low enough but an out-of-equilibrium activity of $^{210}$Pb
at the mBq/kg level was observed, precluding the background goals of
the experiment. Cosmogenic radionuclide production in NaI(Tl) was
also quantified~\cite{ANAIScosmo} and $^{22}$Na and $^{3}$H were
found to be very relevant in the region of interest. A complete
background model of ANAIS-25 data has been developed~\cite{posterp} and the measured background is well understood
down to 3~keVee.

\begin{table}
\centerline{\begin{tabular}{|c|c|c|c|} \hline $^{40}$K (mBq/kg) &
$^{238}$U (mBq/kg) & $^{210}$Pb (mBq/kg) & $^{232}$Th (mBq/kg)
\\\hline
 1.25\,$\pm$\,0.11 (41\,ppb K)& 0.010\,$\pm$\,0.002& 3.15 &0.0020\,$\pm$\,0.0008\\
\hline
\end{tabular}}
\caption{Internal activity measured in the ANAIS-25 detectors.}
\label{tab:cont}
\end{table}

\begin{figure}
 \centerline{\includegraphics[width=0.45\textwidth]{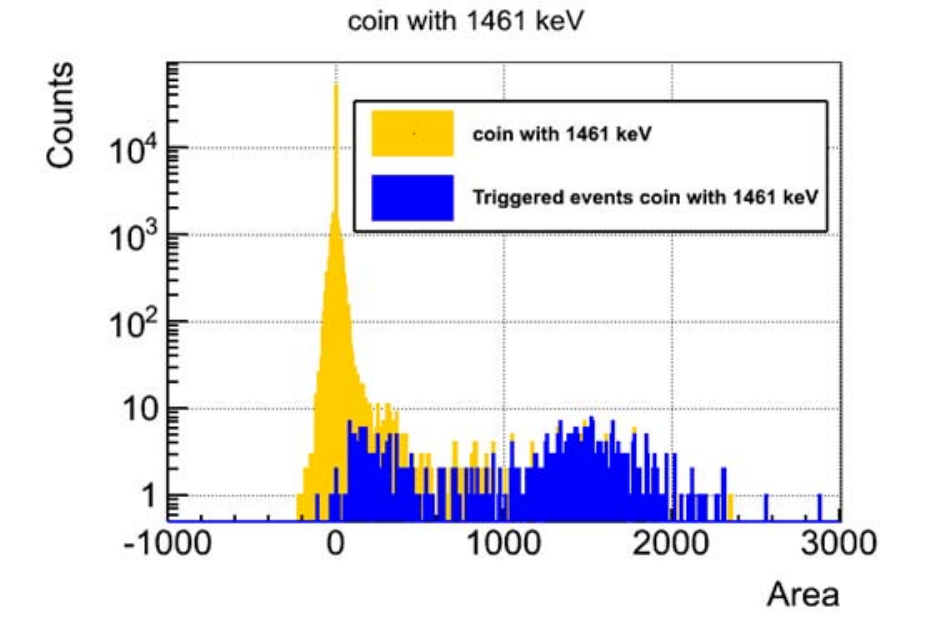}
  \includegraphics[width=0.45\textwidth]{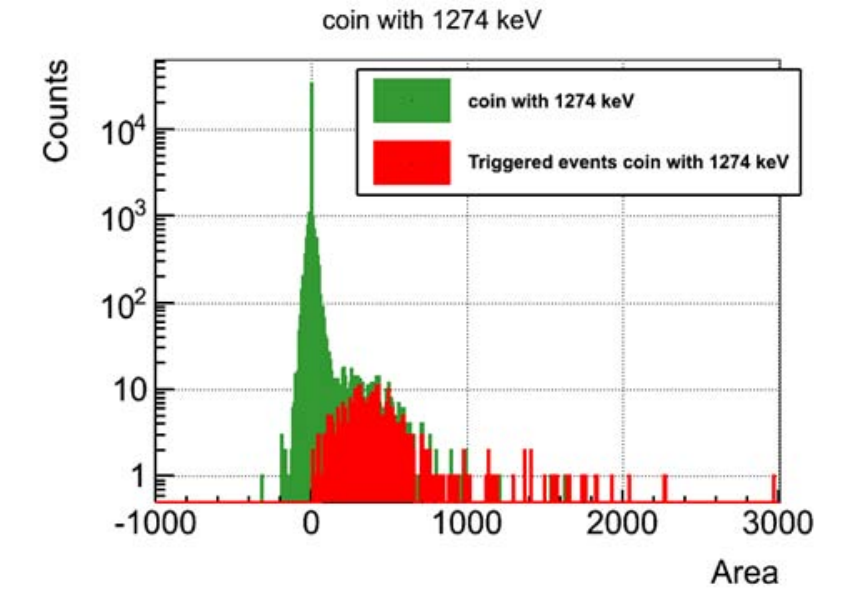}}
  \caption{ANAIS-25 D0 coincident events at low energy for $^{40}$K (left) and for $^{22}$Na (right).}
  \label{fig:A25trigger}
\end{figure}

\begin{figure}
 \centerline{\includegraphics[width=0.45\textwidth]{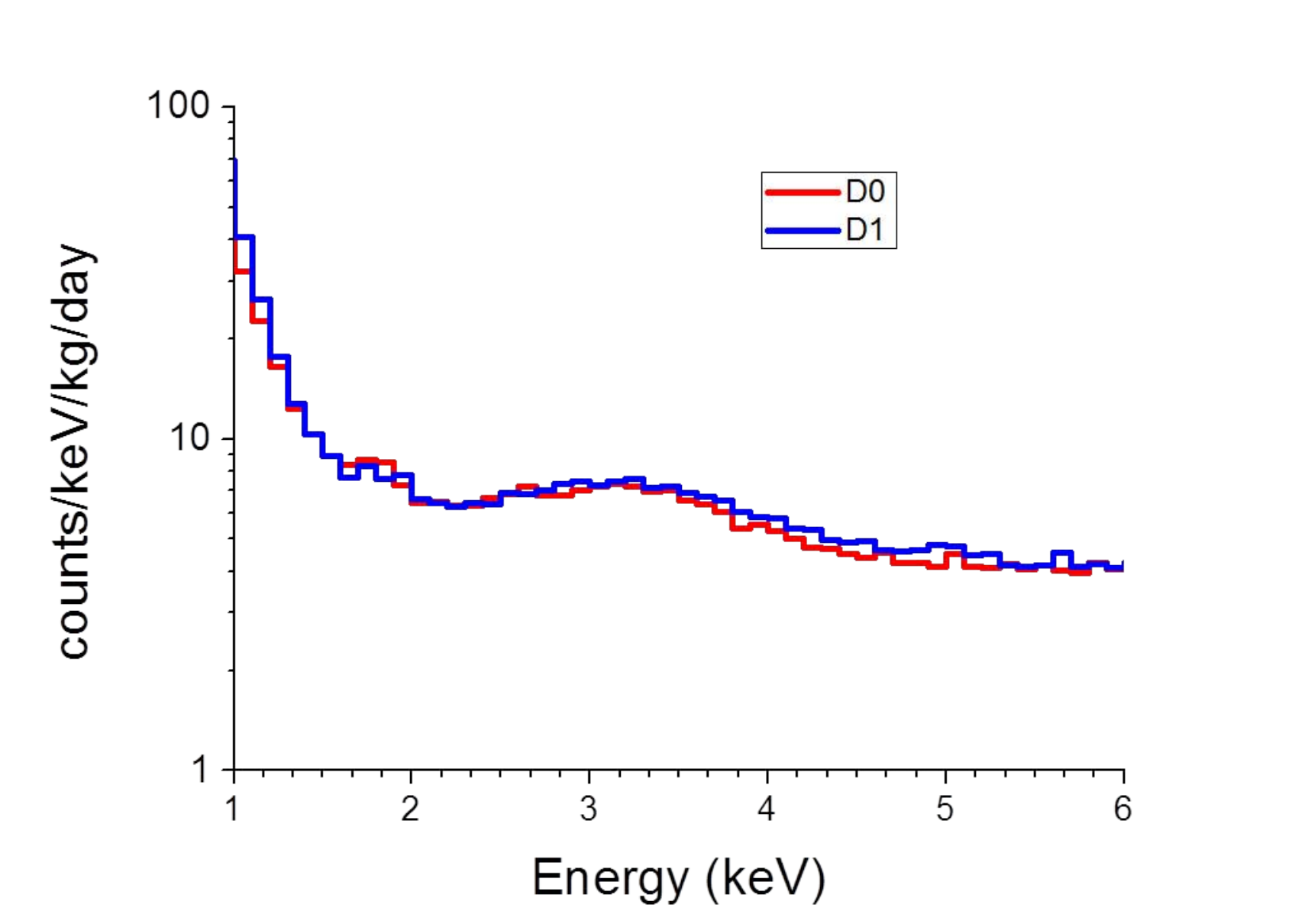}
 \includegraphics[width=0.4\textwidth]{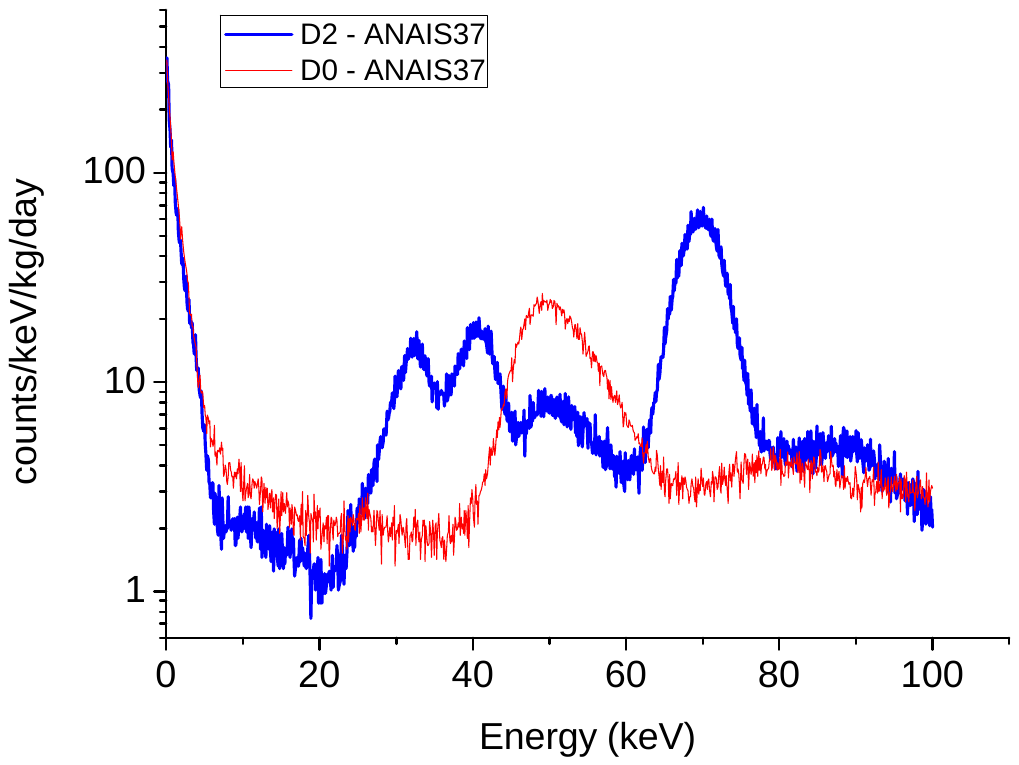}}
  \caption{Left: Preliminary filtered background spectra corrected by triggering and filtering efficiencies for ANAIS-25 detectors D0 and D1 (filtering procedures still being optimized). Right: Raw background spectra of D0 and D2 detectors measured at the ANAIS-37 set-up.}
  \label{fig:A25le}
\end{figure}

The low energy events populations from internal $^{40}$K and $^{22}$Na have been studied. The K-shell electron binding energy following electron capture in $^{40}$K (3.2\,keV) and $^{22}$Na (0.9 keV) can be tagged by the coincidence with a high energy $\gamma$ ray in a second detector (1461\,keV and 1274\,keV respectively). In Fig.~\ref{fig:A25trigger} both populations are shown, together with the events effectively triggering our acquisition; it can be concluded that triggering at 1\,keVee is clearly achieved in ANAIS-25 and therefore an energy threshold of the order of 1\,keVee is at reach. To remove the PMT origin events, dominating the background below 10\,keVee, and then reach the 1\,keVee threshold, specific filtering protocols for ANAIS-25 detectors have been designed following~\cite{ANAISbulk}. A preliminary spectrum, after filtering and correcting by the efficiencies of the cuts, determined with low energy events from a $^{109}$Cd calibration, is shown in Fig.~\ref{fig:A25le}, left.

The origin of the large $^{210}$Pb contamination found in ANAIS-25
crystals was identified and addressed by Alpha Spectra in the
construction of the new module (named D2) integrated in the ANAIS-37
set-up. Very preliminary results corresponding to 50 days of
live-time are presented here for D2. A total alpha rate of
0.58\,$\pm$\,0.01\,mBq/kg has been obtained, which is a factor 5
lower than in D0 and D1, concluding that effective reduction of Rn
entrance in the growing and/or purification at Alpha Spectra has
been achieved. A K content of 44\,$\pm$\,4\,ppb compatible with that
of D0 and D1 (see Table~\ref{tab:cont}) has been measured using the
same technique applied to previous prototypes. The measured light
collection of D2 is compatible with that of ANAIS-25 detectors
too~\cite{posterma} and the measured background of the new
module is well described by the expected
components~\cite{posterp}. Figure~\ref{fig:A25le}, right compares
the raw background spectra of D0 and D2 in the
ANAIS-37 set-up; in spite of the presence of cosmogenic activation in D2 (still decaying) there is a very promising reduction of the background level below 20~keVee.


\section{Sensitivity}
\begin{figure}
\centerline{\includegraphics[width=0.45\textwidth]{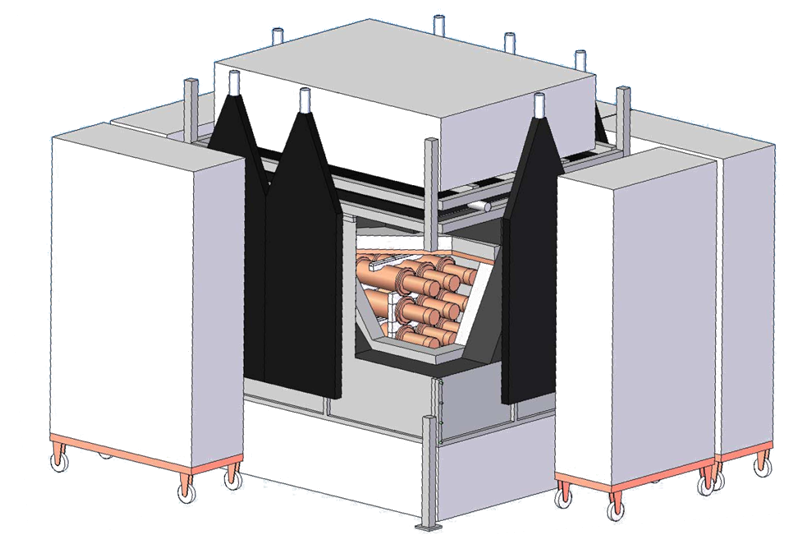}
\includegraphics[width=0.45\textwidth]{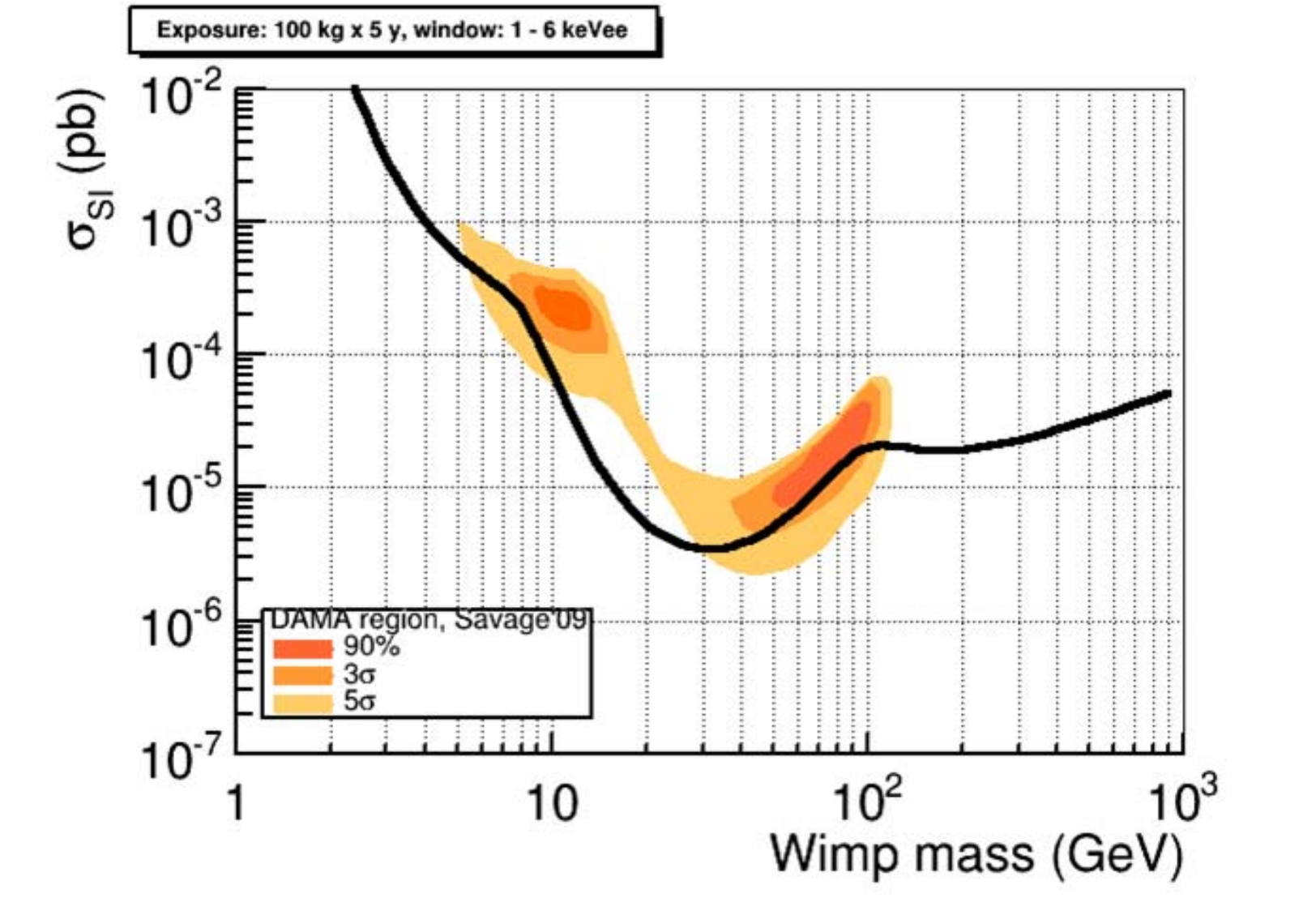}}
  \caption{Left: Design of the ANAIS experiment for a 3$\times$3 crystal matrix with a total mass of 112~kg. Right: Prospects of ANAIS annual modulation sensitivity for 100\,kg total detection mass, presently achieved background without profiting from anticoincidence rejection, five years of data taking and an energy window from 1 to 6\,keVee. These prospects correspond to a detection limit at 90\% CL with a critical limit at 90\% CL, following~\cite{sensitivityPlots}.}
  \label{fig:sens}
\end{figure}

Figure~\ref{fig:sens}, left shows the design for the full ANAIS experiment considering a 3$\times$3 crystal configuration. Prospects of the sensitivity to the annual modulation in the WIMP
mass-cross-section parameter space are shown in
Fig.~\ref{fig:sens}, right for a 100\,kg configuration and 5 years of data
taking. The analysis window considered is from 1 to 6\,keVee. The
background assumed is the one measured in ANAIS-25 (shown in Fig.~\ref{fig:A25le}), but with the
$^{210}$Pb activity measured in the new module D2, i.e. the
contribution of 2.57\,mBq/kg of $^{210}$Pb has been subtracted to
the background measured at ANAIS-25. Further reduction from
anticoincidence measurement, dependent on the detector matrix assumed, is
expected. A conservative approach to derive these prospects
has been followed, but even in this case, there is a considerable
discovery potential of dark matter particles as responsible of the
DAMA/LIBRA signal.

\section*{Acknowledgments}
This work was supported by the Spanish Ministerio de Econom\'{i}a y
Competitividad and the European Regional Development Fund
(MINECO-FEDER) (FPA2011-23749, FPA2014-55986-P), the
Consolider-Ingenio 2010 Programme under grants MULTIDARK
CSD2009-00064 and CPAN CSD2007-00042, and the Gobierno de Arag\'{o}n
(Group in Nuclear and Astroparticle Physics, ARAID Foundation). P.~Villar is
supported by the MINECO Subprograma de Formaci\'{o}n de Personal
Investigador. We also acknowledge LSC and GIFNA staff for their
support.


\begin{footnotesize}

\end{footnotesize}


\end{document}